\documentclass[twocolumn,aps,prl,groupedaddress,amssymb]{revtex4}
\usepackage{natbib}
\usepackage{graphicx}
\usepackage[]{SIunits}
\usepackage{units}
\usepackage{braket}
\usepackage{amsmath}
\usepackage{float}

\begin{document}
\hyphenation{Ryd-berg}

\title{Ultracold atom-electron interaction: from two to many-body physics}

\author{A. Gaj}
\email[Electronic address: ]{a.gaj@physik.uni-stuttgart.de}
\author{A. T. Krupp}
\author{J. B. Balewski}
\author{R. L\"{o}w}
\author{S. Hofferberth}
\affiliation{5. Physikalisches Institut, Universit\"{a}t
Stuttgart, Pfaffenwaldring 57, 70569 Stuttgart, Germany}
\author{T. Pfau}
\email[Electronic address: ]{t.pfau@physik.uni-stuttgart.de}
\affiliation{5. Physikalisches Institut, Universit\"{a}t
Stuttgart, Pfaffenwaldring 57, 70569 Stuttgart, Germany}
\date{15 March 2014}
\begin{abstract}

\end{abstract}
\maketitle

The transition from a few-body system to a many-body system can result in new 
length scales, novel collective phenomena or even in a phase transition. Such a 
threshold behavior was shown for example in $^4$He droplets, where $^4$He turns 
into a superfluid for a  specific number of particles \cite{Hartman1996}. A 
particularly interesting question in this context is at which point a few-body 
theory can be substituted by a mean field model, i.\,e. where the discrete number 
of particles can be treated as a continuous quantity. Such a transition from two 
non-interacting fermionic particles to a Fermi sea was demonstrated recently 
\cite{Jochim2013}. In this letter, we study a similar crossover to a many-body 
regime based on ultralong-range Rydberg molecules \cite{Greene2000} forming a 
model system with binary interactions. This class of exotic molecules shows very 
weak binding energies, similar to magneto-associated Feshbach molecules \cite{Herbig2003,Koehler2006}, 
and thus requires ultracold temperatures. A wide range of fascinating phenomena, 
starting from the coherent creation and breaking of chemical bonds \cite{Butscher2010} 
to a permanent electric dipole moment in a homonuclear molecule 
\cite{Li2011}, has been studied. Dimers, consisting of a single atom in the 
Rydberg state and one atom in the ground state, have been observed in an 
ultracold gas of Rb in the Rydberg \textit{S}-state 
\cite{Bendkowsky2009}, \textit{D}-state \cite{Krupp2014,Anderson2014} and \textit{P}-state 
\cite{Bellos2013}, and of Cs \cite{Tallant2012} in the Rydberg \textit{S}-state. The many-body 
regime, where the Rydberg electron experiences a mean shift by thousands of atoms 
within its orbit, has been studied at very high densities in a BEC, leading to 
electron-phonon coupling \cite{Balewski2013}, and in a hot vapor \cite{Amaldi1934}. 
Here, we trace the transition between the two regimes in an ultracold cloud with 
a constant density by tuning only one parameter: the principal quantum number 
\textit{n} of the excited Rydberg state. We probe the border of the mean field 
limit, where the energy and length scales of these molecules become extreme. 
While the binding energies are the smallest ever observed for this type of 
molecules, the size reaches the dimensions of macroscopic biological objects 
like viruses or bacteria. \\ \indent
The bond in ultralong-range Rydberg molecules results from the scattering of a
slow Rydberg electron from a neutral atom with a negative scattering length 
\textit{a} \cite{Greene2000}. The theoretical approach is based on Fermi's 
original concept of the pseudopotential \cite{Fermi1934}
\begin{equation}
V_{\text{pseudo}}(\textbf{r},\textbf{R})=\frac{2\pi \hbar^2 a}{m_e} \delta (\textbf{r}-\textbf{R})
\label{eq:1}
\end{equation}
describing such a scattering event between an electron of mass $m_e$ at position 
\textbf{r} and an atom  at \textbf{R}. If the scattering 
length \textit{a} is negative, the interaction is attractive and ground state 
atoms can be bound in a potential
\begin{equation}
V(\textbf{R})=\frac{2 \pi \hbar^2 a}{m_e} \left| \Psi (\textbf{R}) \right|^2,
\label{eq:2}
\end{equation}
where $\left| \Psi (\textbf{R}) \right|^2$ is the local electron density. 
In case of $^{87}$Rb the theoretical value of the scattering length \textit{a} 
for triplet scattering is $-$\unit[16.1]{$a_0$} \cite{Bahrim2001}, where $a_0$ 
is the Bohr radius. The singlet scattering length is positive and therefore does 
not lead to a bound state. Momentum-dependent corrections to these values can be 
estimated using a semiclassical approximation \cite{Omont1977}. 
\begin{figure}[!b]
\includegraphics[width=87mm]{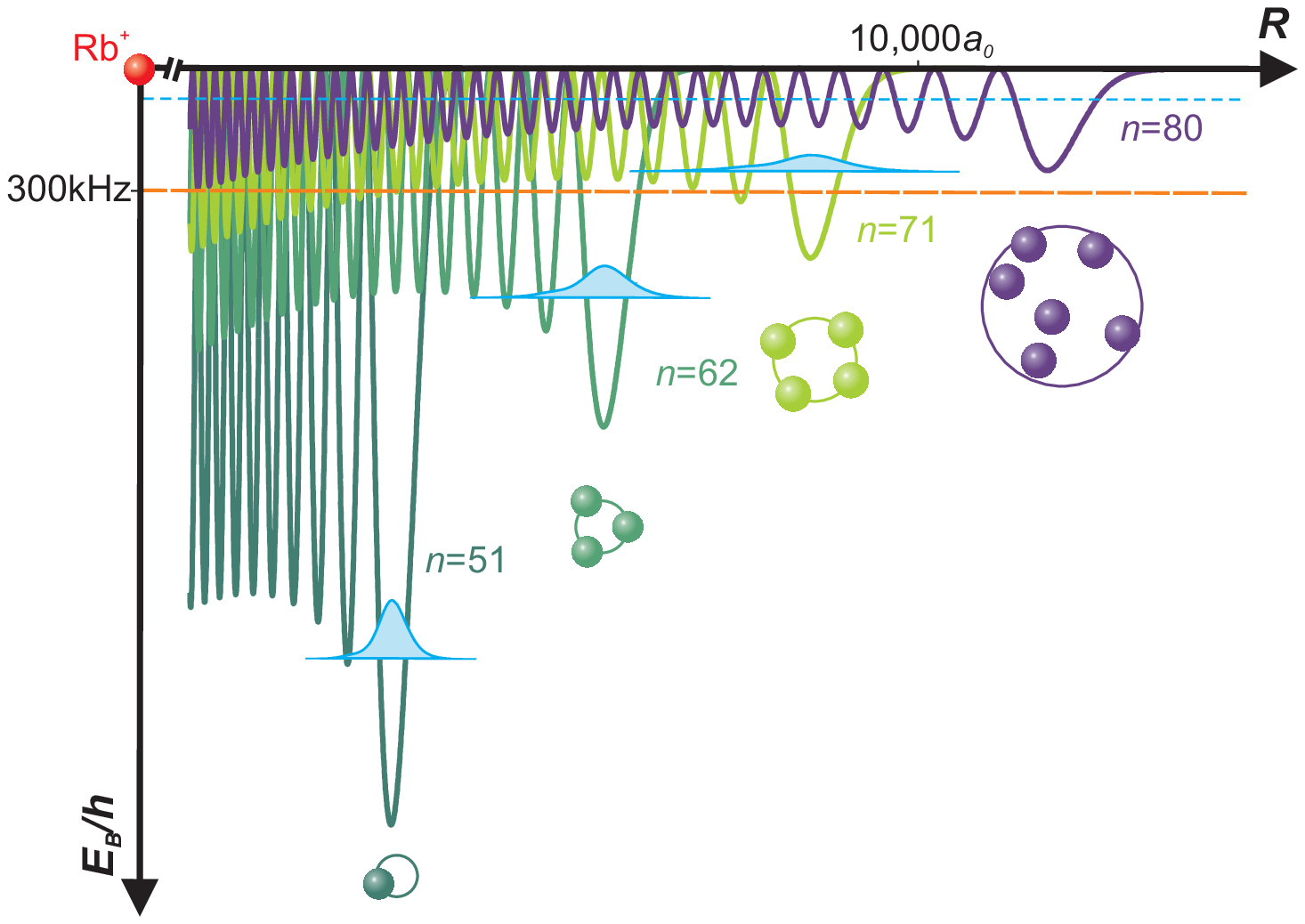}
\caption{ \textbf{Scattering potentials calculated from equation (\ref{eq:2}) for \textit{n}=51,\,62,\,71 
and 80.} Dimers are predominantly bound in the vibrational ground state in the 
outermost well (light blue). For \textit{n}=80 a single molecular state localized 
in the outermost lobe cannot be resolved in the experiment, because the mean binding 
energy per atom becomes too low (light blue dashed line) and thus a change from 
a few-body to a many-body description is required. The spheres below illustrate 
the highest order poly-mer observed for each principal quantum number. Orange 
dashed line indicates the mean shift for the experimental density.}
\label{Fig1}
\end{figure}
Additionally, a \textit{p}-wave shape resonance can cause a substantial modification of 
the molecular potential, leading to butterfly-type molecules \cite{Hamilton2002} 
and states bound by internal quantum reflection \cite{Bendkowsky2010}. These 
corrections are largely negligible for high principal quantum numbers and large 
distances from the Rydberg core, where the relative motion of the Rydberg electron 
and the perturber is slow. The ground state wavefunction of the Rydberg molecules 
with the highest binding energy per atom, is localized mainly at the position of 
the outermost lobe of the electron wavefunction, i.\,e. near the classical turning 
point of the electron (see Fig.\,\ref{Fig1}). \\ \indent
Molecules containing more than one ground state atom can be described with the 
same formalism. Two ground state atoms inside the electron orbit essentially do not interact 
with each other, since the Rb atom-atom scattering length \cite{Julienne1997} 
is much smaller than the mean particle distance in a dilute thermal cloud. Hence, the 
binding energy of an \textit{i}-atomic molecule, \textit{i}\,$\in\mathbb{N}$, 
is $(i-1)$ times larger than the binding energy of a dimer. Only when the number 
of ground state atoms inside the Rydberg atom becomes very large, a change in the 
description of the system from discrete bound states to a mean field approximation 
is required (Fig.\,\ref{Fig1}). In this case no individual bound states are 
resolved, but the Rydberg line is shifted by
\begin{equation}
\begin{split}
\Delta E&=\iint d\textbf{r} d\textbf{R}  \ V_{\text{pseudo}}(\textbf{r},\textbf{R}) \left| \Psi (\textbf{r}) \right|^2 \rho (\textbf{R})\\
        &=\int d\textbf{R} \ V(\textbf{R}) \rho(\textbf{R}) = \frac{2 \pi \hbar^2 a}{m_e} \bar{\rho}.
\end{split}
\label{eq:3}
\end{equation}
If higher order corrections to the zero-energy scattering length can be neglected,
the mean shift depends only on the value of \textit{a} and the density 
$\bar{\rho}$ averaged over the volume of the Rydberg atom. \\ \indent
We perform the experiment in a magnetically trapped ultracold cloud of $^{87}$Rb 
atoms in the 5\textit{S}$_{1/2}$, F=2, m$_F$=2 ground state with typical temperatures of 
\unit[2]{\micro K} and densities on the order of \unit[10$^{12}$]{cm$^{-3}$}. 
Detailed information about the setup can be found in \cite{LWN12}. After the 
preparation of the ultracold cloud we excite the atoms in a two photon process 
via the $6P_{3/2}$ state to the $nS_{1/2}$ Rydberg state. For our high precision 
spectroscopy measurements we use narrow bandwidth lasers ($\le$\unit[30]{kHz}), 
which are locked to a high finesse ULE reference cavity. The \unit[420]{nm} light, 
driving the lower transition, is blue detuned from the intermediate state by 
\unit[80]{MHz} to avoid absorption and heating of the cloud. It is sent to the 
experiment in \unit[50]{\micro s} pulses with a repetition rate of \unit[167]{Hz}. 
During the sequence the \unit[1016]{nm} laser light driving the upper 
transition is on constantly. After the excitation we field-ionize the Rydberg atoms and 
collect the ions on a microchannel plate detector. In a single atomic cloud we perform 
typically 400 cycles of Rydberg excitation and detection while scanning the 
frequency of the blue laser light. In order to realize high spectral resolution 
we choose long excitation pulses of \unit[50]{\micro s}. Taking further into account 
the laser bandwidth, Doppler broadening and natural linewidth this results in an 
experimental resolution of around \unit[60]{kHz}. In order to obtain the best 
visibility while changing the principal quantum number of the excited Rydberg state, 
we adjust the power of the blue laser to account for power broadening. Only the 
spectrum of \textit{n}=51 was taken with higher laser power and thus in this case 
the atomic line is slightly broadened. 
\begin{figure}[H]
\includegraphics[width=88mm]{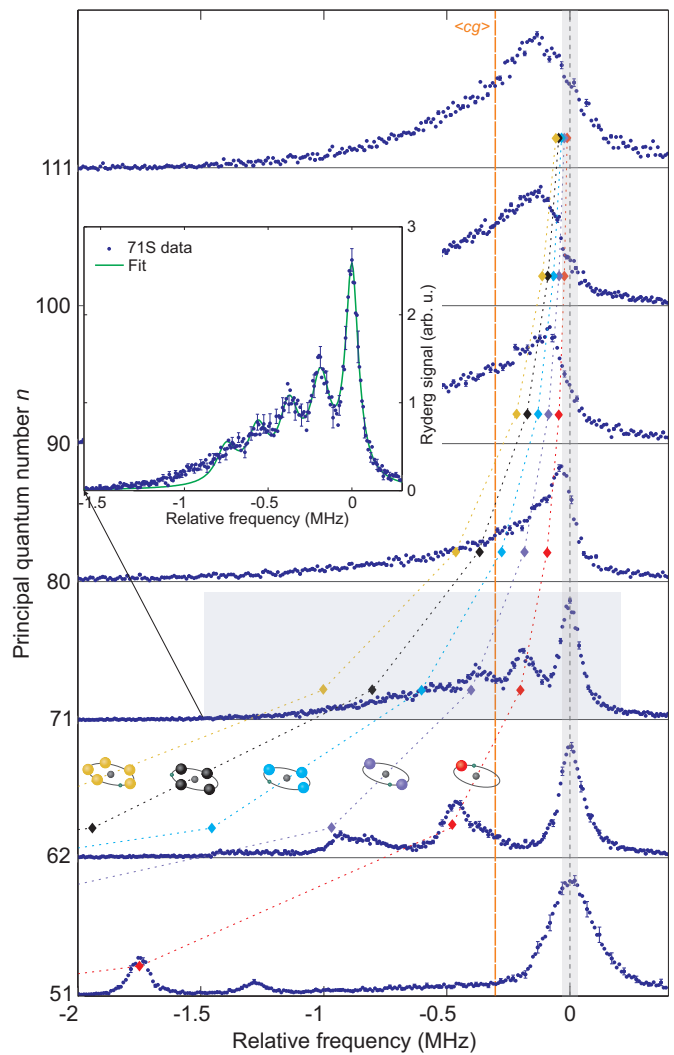}
\caption{ \textbf{Overview of the $5S_{1/2}$ to $nS_{1/2}$ excitation spectra showing 
the dimer - poly-mer transition for increasing principal quantum number.} The origin of the 
relative frequency axis corresponds to the center of the atomic line (dashed line) for 
\textit{n}$\le$71, where molecular lines are distinguishable. 
The gray shaded area between $-$\unit[30]{kHz} and \unit[30]{kHz} indicates 
the laser bandwidth. Spectra at \textit{n}$>$71 are 
horizontally shifted such that their centers of gravity overlap with the 
mean center of gravity $\left\langle cg\right\rangle$ (orange dashed line) of the first 
three spectra (\textit{n}=51,\,62,\,71). All data was taken at similar cloud 
parameters, therefore the density induced shift for all spectra is constant to first 
approximation. Molecules with up to three bound ground state atoms for 62\textit{S} and 
up to four for 71\textit{S} are resolvable in the spectra. Colored diamonds indicate the 
positions of the dimers (red), trimers (violet) etc. following the power law 
scaling of the binding energies fitted to the first three spectra. In the inset 
the molecular spectrum for the \textit{n}=71 Rydberg state is shown. A 
multilorentzian fit (green line), assuming a constant spacing between the molecular 
peaks is plotted to indicate the positions of the higher order molecular lines. 
The spectrum for \textit{n}=40 is not shown, because the binding energy of the dimer 
is larger than the plotting range. Each spectrum is an average over 20 independent 
measurements with standard deviation error bars.} 
\label{Fig2} \end{figure}
In Fig.\,\ref{Fig2} excitation spectra from the $5S_{1/2}$ to the $nS_{1/2}$ state, 
where \textit{n} is ranging from 51 to 111, are presented.
The shape of the obtained 
Rydberg spectra varies significantly for different \textit{n}. 
For low principal quantum numbers clearly distinguishable molecular lines are 
present on the red side of the atomic peak, which is situated at the origin. 
In the spectrum of \textit{n}=51 the peak at $-$\unit[1.7]{MHz} can be identified 
as a dimer, for which the ground state atom is localized in the outermost well 
of the molecular potential. Additionally at $-$\unit[1.3]{MHz} an excited vibrational 
state \cite{Bendkowsky2009,Bendkowsky2010} is visible. For a given density, the 
probability to find an atom inside the electron orbit scales as ${\textit{n}^*}^{6}$ 
with the effective principal quantum number $n^*=n-\delta$, where $\delta$ is 
the quantum defect. 
Hence, higher order molecules are formed more likely at higher \textit{n}. 
At the same time the binding energy per atom decreases and thus polyatomic 
molecular lines become visible in the spectra of \textit{n}=62 and 71. The binding 
energy \textit{E}$_B$ can be directly measured as the difference between the atomic and the
molecular line in the spectrum. At \textit{n}=62 lines up to the tetramer together 
with corresponding vibrational excited states are visible. The broadening of the 
tetramer line may be caused by the presence of these excited states,
possibly with a reduced lifetime \cite{Butscher2011}. At \textit{n}=71 only 
vibrational ground states are resolved. Polyatomic molecules up to a pentamer 
can be identified. The size of such a molecule becomes enormous due to the Rydberg 
electron orbit radius reaching almost 10\,000$a_0$. For large \textit{n} the binding 
energy $E_B$ decreases until it is below the experimental resolution. This 
manifests in a non-resolvable shoulder and finally in an inhomogeneously broadened
spectral line. \\ \indent
The experimental binding energies can be calculated based on the molecular 
potential (\ref{eq:2}) and the mean shift (\ref{eq:3}). We use corrections to the 
\textit{s}-wave scattering length including terms linear in the relative momentum of two 
scattering partners based on a semiclassical approximation \cite{Omont1977}. 
This approximation is valid for large distances from the ionic core. Therefore we 
restrict the analysis only to the lowest bound state. The discussion of higher 
vibrational states can be found in \cite{Bendkowsky2010}. We solve the Schr\"{o}dinger 
equation for the ground state atom in the molecular potential using Numerov's 
method and fit the zero-energy scattering length \textit{a} to the binding energies 
of the dimers. In this paper the best agreement with the experimental data is 
obtained for $-$\unit[16.2]{$a_0$}, which is very close to the theoretically 
predicted value of $-$\unit[16.1]{$a_0$} \cite{Bahrim2001}. \\ \indent
\begin{figure}[!t]
\includegraphics[width=87mm]{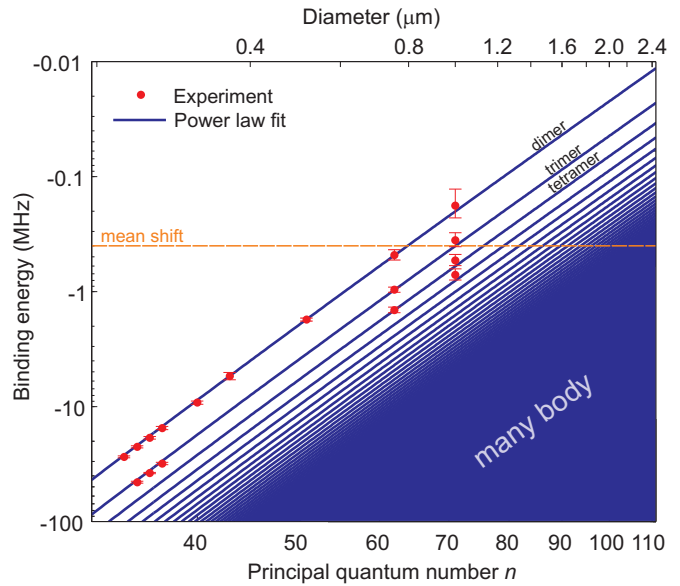}
\caption{ \textbf{Measured binding energies (red points) versus principal quantum number 
(bottom axis) and diameters (top axis) of the molecules.} The data for \textit{n}$\le$37 
is taken from \cite{Bendkowsky2009}. For \textit{n}=40,\,43,\,51 the frequency range 
chosen in the experiment was too small to photoassociate molecules with larger 
binding energies than a dimer. The power law ${\textit{n}^*}^{b}$ (blue lines) 
fitted to the measured data and multiplied by factors $i-1$, $\textit{i}\,\in \mathbb{N}$, 
(\textit{i}=2 for a dimer) shows that for \textit{n}$>$75 the binding energy of 
the dimer becomes smaller than the experimental resolution. The increasing number 
of ground state atoms inside the electron orbit leads eventually to a mean shift 
of the Rydberg line. Calculated values of \textit{E}$_B$ are not shown in the plot, 
because they are hardly distinguishable from the experimental data on this scale. 
The error bars are determined as the standard deviation of the fit. }
\label{Fig3}
\end{figure} 
For \textit{n}$>$71, where no distinct molecular lines can be identified, a mean
field description is required. Furthermore, the spectral position of pure 
Rydberg atoms, and thus the zero position, cannot be identified directly from the signal. 
However it can be determined from the center of gravity $cg$ of the spectra, 
taken in a non-blockaded sample, which is constant for a given density. Intuitively, 
this result in the first approximation can be explained by the fact, that while 
the mean potential depth $\bar{V}=\frac{\int d\textbf{R} \ V(\textbf{R})}{\int d\textbf{R}}$ 
averaged over the volume of the Rydberg atom decreases with the 
effective principal number as ${\textit{n}^*}^{-6}$, the probability to find an 
atom inside the Rydberg electron orbit increases with ${\textit{n}^*}^{6}$. 
In the experiment all data was taken at a fixed density. Thus, in our analysis we overlap the center of 
gravity of the spectra at \textit{n}$\ge$71 with the mean $\left\langle cg\right\rangle$=\unit[$-300$]{kHz} 
determined at low principal quantum numbers. Doing so we can identify the zero 
position in the top panels of Fig.\,\ref{Fig2}. Assuming the scattering length 
\textit{a} to be constant, we determine the effective density to be 
\unit[3$\cdot$10$^{12}$]{cm$^{-3}$}, which is close to the peak density obtained 
from a Gaussian fit to absorption images of the thermal cloud. This indicates 
that molecules are most likely created in regions of high density. Only for 
\textit{n}$\le$80 the highest signal originates from pure Rydberg atoms. 
Already for \textit{n}=80 there is on average one ground state atom inside the 
electron orbit, leading to a high probability to excite dimers instead of pure 
Rydberg atoms. On average there are four atoms inside the 100S electron orbit 
and eight for the 111S state.
Therefore, the atomic line is suppressed, while the molecular lines 
are not resolvable in the experiment any more, caused by their very low binding 
energies. According to the central limit theorem, for even higher principal quantum 
numbers the shape of the spectrum is expected to become Gaussian, with the maximum at the 
position of the mean shift ($\left\langle cg\right\rangle$ in Fig.\,\ref{Fig2}). 
The density dependent dephasing resulting from the existence of many molecular lines 
within the Rydberg line envelope sets a fundamental limit for the number of atoms 
inside the blockade radius i.\,e. the optical thickness. This fact is of importance for 
every experiment taken at high principal quantum numbers and high densities, in particular 
for quantum optics experiments in ultracold clouds \cite{Peyronel2012,Dudin2012,Baur2014}.\\ \indent
The binding energies of all observed molecules are plotted in Fig.\,\ref{Fig3}. 
From the extrapolated molecular binding energies the transition from the few 
particle description of discrete bound states to a mean field shift becomes 
visible. A power law fitted to the \textit{n}$\ge$40 data shows a scaling with 
the effective principal quantum number \textit{n}$^*$ to the power of $-6.26\pm 0.12$, 
close to the value of $-6$, expected from the size scaling argument. The deviation 
can be explained by the dependence of the scattering length on the relative momentum 
and the fact, that with increasing principal quantum number the shape of the outermost 
well of the molecular potential changes. Taking the corresponding zero-point energy and momentum 
dependent corrections to the scattering length into account, we obtain an exponent 
of $-6.37$, which is in very good agreement with our experimental data. Contributions 
of higher order partial waves and \textit{p}-wave shape resonances to the molecular 
potential can be neglected since the kinetic energy associated with the relative 
motion of two scattering partners is small in the region of interest. \\ \indent
Rydberg molecules in an ultracold cloud constitute a tunable model system to study 
the transition from a few-body to a many-body regime. They offer a unique tool to 
address few-body subsystems with control on a single particle level by changing 
the detuning of the excitation laser light. Complementary to this work the number 
of constituents and the interaction strength can be also varied independently by 
changing the density and the principal quantum number of the excited Rydberg atom. 
Furthermore, the analysis of the relative strength of the molecular lines opens up 
the possibility to measure correlations in a bosonic gas. In high density gases 
and for low principal quantum numbers, where the size of the Rydberg atom is comparable 
to the de Broglie wavelength, extracting the \textit{g}$^{(2)}$ correlation function 
\cite{Naraschewski1999} of thermal and Bose-condensed gases is feasible. In addition to 
previous measurements \cite{Burt1997,Hodgman2011} also higher order correlation 
functions can be studied using polyatomic molecules.

\begin{acknowledgments}
We acknowledge support from Deutsche Forschungsgemeinschaft (DFG) within the 
SFB/TRR21 and the project PF 381/4-2. Parts of this work was also founded by
ERC under contract number 267100. A.G. acknowledges support from E.U. Marie Curie 
program ITN-Coherence 265031 and S.H. from DFG through the project HO 4787/1-1. 
\end{acknowledgments}

\noindent 
The experiment was conceived by A.G., A.T.K., R.L., S.H. and T.P. and carried out 
by A.G. and A.T.K. ; data analysis was accomplished A.G. and A.T.K.; numerical 
calculation is by A.G. and J.B.B. and A.G. wrote the manuscript with contributions 
from all authors.

\noindent 
The authors declare that they have no competing financial interests.


\begin{thebibliography}{10}
\expandafter\ifx\csname url\endcsname\relax
  \def\url#1{\texttt{#1}}\fi
\expandafter\ifx\csname urlprefix\endcsname\relax\def\urlprefix{URL }\fi
\providecommand{\bibinfo}[2]{#2}
\providecommand{\eprint}[2][]{\url{#2}}

\bibitem{Hartman1996}
\bibinfo{author}{Hartman, M.}, \bibinfo{author}{Miller, R.~E.},
  \bibinfo{author}{Toennies, J.~P.} \& \bibinfo{author}{Vilesov, A.~F.}
\newblock \bibinfo{title}{High-resolution molecular spectroscopy of van der
  {W}aals clusters in liquid helium droplets}.
\newblock \emph{\bibinfo{journal}{Science}} \textbf{\bibinfo{volume}{272}},
  \bibinfo{pages}{1631--1634} (\bibinfo{year}{1996}).

\bibitem{Jochim2013}
\bibinfo{author}{Wenz, A.} \emph{et~al.}
\newblock \bibinfo{title}{From few to many: Observing the formation of a {F}ermi
  sea one atom at a time}.
\newblock \emph{\bibinfo{journal}{Science}} \textbf{\bibinfo{volume}{342}},
  \bibinfo{pages}{457--460} (\bibinfo{year}{2013}).

\bibitem{Greene2000}
\bibinfo{author}{Greene, C.~H.}, \bibinfo{author}{Dickinson, A.~S.} \&
  \bibinfo{author}{Sadeghpour, H.~R.}
\newblock \bibinfo{title}{Creation of polar and nonpolar ultra-long-range
  {R}ydberg molecules}.
\newblock \emph{\bibinfo{journal}{Phys. Rev. Lett.}}
  \textbf{\bibinfo{volume}{85}}, \bibinfo{pages}{2458--2461}
  (\bibinfo{year}{2000}).

\bibitem{Herbig2003}
\bibinfo{author}{Herbig, J.} \emph{et~al.}
\newblock \bibinfo{title}{Preparation of a pure molecular quantum gas}.
\newblock \emph{\bibinfo{journal}{Science}} \textbf{\bibinfo{volume}{301}},
  \bibinfo{pages}{1510--1513} (\bibinfo{year}{2003}).

\bibitem{Koehler2006}
\bibinfo{author}{K\"ohler, T.}, \bibinfo{author}{G\'oral, K.} \&
  \bibinfo{author}{Julienne, P.~S.}
\newblock \bibinfo{title}{Production of cold molecules via magnetically tunable
  {F}eshbach resonances}.
\newblock \emph{\bibinfo{journal}{Rev. Mod. Phys.}}
  \textbf{\bibinfo{volume}{78}} (\bibinfo{year}{2006}).

\bibitem{Butscher2010}
\bibinfo{author}{Butscher, B.} \emph{et~al.}
\newblock \bibinfo{title}{Atom-molecule coherence for ultralong-range {R}ydberg
  dimers}.
\newblock \emph{\bibinfo{journal}{Nature Phys}} \textbf{\bibinfo{volume}{6}},
  \bibinfo{pages}{1745--2473} (\bibinfo{year}{2010}).

\bibitem{Li2011}
\bibinfo{author}{Li, W.} \emph{et~al.}
\newblock \bibinfo{title}{A homonuclear molecule with a permanent electric
  dipole moment}.
\newblock \emph{\bibinfo{journal}{Science}} \textbf{\bibinfo{volume}{334}},
  \bibinfo{pages}{1110--1114} (\bibinfo{year}{2011}).

\bibitem{Bendkowsky2009}
\bibinfo{author}{Bendkowsky, V.} \emph{et~al.}
\newblock \bibinfo{title}{Observation of ultralong-range {R}ydberg molecules}.
\newblock \emph{\bibinfo{journal}{Nature}} \textbf{\bibinfo{volume}{458}},
  \bibinfo{pages}{0028--0836} (\bibinfo{year}{2009}).

\bibitem{Krupp2014}
\bibinfo{author}{Krupp, A.~T.} \emph{et~al.}
\newblock \bibinfo{title}{Alignment of {D}-state {R}ydberg molecules}.
\newblock \emph{\bibinfo{journal}{Phys. Rev. Lett.}}
  \textbf{\bibinfo{volume}{112}}, \bibinfo{pages}{143008}
  (\bibinfo{year}{2014}).

\bibitem{Anderson2014}
\bibinfo{author}{Anderson, D.~A.}, \bibinfo{author}{Miller, S.~A.} \&
  \bibinfo{author}{Raithel, G.}
\newblock \bibinfo{title}{Photoassociation of long range $nd$ {R}ydberg
  molecules}.
\newblock \emph{\bibinfo{journal}{arXiv:1401.2477}}  (\bibinfo{year}{2014}).

\bibitem{Bellos2013}
\bibinfo{author}{Bellos, M.~A.} \emph{et~al.}
\newblock \bibinfo{title}{Excitation of weakly bound molecules to trilobitelike
  {R}ydberg states}.
\newblock \emph{\bibinfo{journal}{Phys. Rev. Lett.}}
  \textbf{\bibinfo{volume}{111}}, \bibinfo{pages}{053001}
  (\bibinfo{year}{2013}).

\bibitem{Tallant2012}
\bibinfo{author}{Tallant, J.}, \bibinfo{author}{Rittenhouse, S.~T.},
  \bibinfo{author}{Booth, D.}, \bibinfo{author}{Sadeghpour, H.~R.} \&
  \bibinfo{author}{Shaffer, J.~P.}
\newblock \bibinfo{title}{Observation of blueshifted ultralong-range {C}s$_2$
  {R}ydberg molecules}.
\newblock \emph{\bibinfo{journal}{Phys. Rev. Lett.}}
  \textbf{\bibinfo{volume}{109}}, \bibinfo{pages}{173202}
  (\bibinfo{year}{2012}).

\bibitem{Balewski2013}
\bibinfo{author}{Balewski, J.~B.} \emph{et~al.}
\newblock \bibinfo{title}{Coupling a single electron to a {B}ose-{E}instein
  condensate}.
\newblock \emph{\bibinfo{journal}{Nature}} \textbf{\bibinfo{volume}{502}},
  \bibinfo{pages}{664--667} (\bibinfo{year}{2013}).

\bibitem{Amaldi1934}
\bibinfo{author}{Amaldi, E.} \& \bibinfo{author}{Segr\`{e}, E.}
\newblock \bibinfo{title}{Effect of pressure on high terms of alkaline
  spectra}.
\newblock \emph{\bibinfo{journal}{Nature}} \textbf{\bibinfo{volume}{133}},
  \bibinfo{pages}{141} (\bibinfo{year}{1934}).

\bibitem{Fermi1934}
\bibinfo{author}{Fermi, E.}
\newblock \bibinfo{title}{Sopra lo spostamento per pressione delle righe
  elevate delle serie spettrali}.
\newblock \emph{\bibinfo{journal}{Nuovo Cimento}}
  \textbf{\bibinfo{volume}{11}}, \bibinfo{pages}{157--166}
  (\bibinfo{year}{1934}).

\bibitem{Bahrim2001}
\bibinfo{author}{Bahrim, C.}, \bibinfo{author}{Thumm, U.} \&
  \bibinfo{author}{Fabrikant, I.~I.}
\newblock \bibinfo{title}{$^3\text{S}_e$ and $^1\text{S}_e$ scattering lengths
  for $e^-$ + {R}b, {C}s and {F}r collisions}.
\newblock \emph{\bibinfo{journal}{J. Phys. B: At. Mol. Opt. Phys.}}
  \textbf{\bibinfo{volume}{34}}, \bibinfo{pages}{L195--L201}
  (\bibinfo{year}{2001}).

\bibitem{Omont1977}
\bibinfo{author}{Omont, A.}
\newblock \bibinfo{title}{On the theory of collisions of atoms in {R}ydberg
  states with neutral particles}.
\newblock \emph{\bibinfo{journal}{J. Phys. France}}
  \textbf{\bibinfo{volume}{38}}, \bibinfo{pages}{1343--1359}
  (\bibinfo{year}{1977}).

\bibitem{Hamilton2002}
\bibinfo{author}{Hamilton, E.~L.}, \bibinfo{author}{Greene, C.~H.} \&
  \bibinfo{author}{Sadeghpour, H.~R.}
\newblock \bibinfo{title}{Shape-resonance-induced long-range molecular {R}ydberg
  states}.
\newblock \emph{\bibinfo{journal}{J. Phys. B: At. Mol. Opt. Phys.}}
  \textbf{\bibinfo{volume}{35}} (\bibinfo{year}{2002}).

\bibitem{Bendkowsky2010}
\bibinfo{author}{Bendkowsky, V.} \emph{et~al.}
\newblock \bibinfo{title}{Rydberg trimers and excited dimers bound by internal
  quantum reflection}.
\newblock \emph{\bibinfo{journal}{Phys. Rev. Lett.}}
  \textbf{\bibinfo{volume}{105}}, \bibinfo{pages}{163201}
  (\bibinfo{year}{2010}).

\bibitem{Julienne1997}
\bibinfo{author}{Julienne, P.~S.}, \bibinfo{author}{Mies, F.~H.},
  \bibinfo{author}{Tiesinga, E.} \& \bibinfo{author}{Williams, C.~J.}
\newblock \bibinfo{title}{Collisional stability of double {B}ose condensates}.
\newblock \emph{\bibinfo{journal}{Phys. Rev. Lett.}}
  \textbf{\bibinfo{volume}{78}}, \bibinfo{pages}{1880} (\bibinfo{year}{1997}).

\bibitem{LWN12}
\bibinfo{author}{L\"ow, R.} \emph{et~al.}
\newblock \bibinfo{title}{An experimental and theoretical guide to strongly
  interacting {R}ydberg gases}.
\newblock \emph{\bibinfo{journal}{J.~Phys.~B: At.~Mol.~Opt.~Phys.}}
  \textbf{\bibinfo{volume}{45}}, \bibinfo{pages}{113001}
  (\bibinfo{year}{2012}).

\bibitem{Butscher2011}
\bibinfo{author}{Butscher, B.} \emph{et~al.}
\newblock \bibinfo{title}{Lifetimes of ultralong-range {R}ydberg molecules in
  vibrational ground and excited states}.
\newblock \emph{\bibinfo{journal}{J. Phys. B: At. Mol. Opt. Phys.}}
  \textbf{\bibinfo{volume}{44}}, \bibinfo{pages}{184004}
  (\bibinfo{year}{2011}).

\bibitem{Peyronel2012}
\bibinfo{author}{Peyronel, T.} \emph{et~al.}
\newblock \bibinfo{title}{Quantum nonlinear optics with single photons enabled
  by strongly interacting atoms}.
\newblock \emph{\bibinfo{journal}{Nature}} \textbf{\bibinfo{volume}{488}},
  \bibinfo{pages}{57--60} (\bibinfo{year}{2012}).

\bibitem{Dudin2012}
\bibinfo{author}{Dudin, Y.~O.}, \bibinfo{author}{Li, L.},
  \bibinfo{author}{Bariani, F.} \& \bibinfo{author}{Kuzmich, A.}
\newblock \bibinfo{title}{Observation of coherent many-body {R}abi oscillations}.
\newblock \emph{\bibinfo{journal}{Nature Phys}} \textbf{\bibinfo{volume}{8}}
  (\bibinfo{year}{2012}).

\bibitem{Baur2014}
\bibinfo{author}{Baur, S.}, \bibinfo{author}{Tiarks, D.},
  \bibinfo{author}{Rempe, G.} \& \bibinfo{author}{D\"{u}rr, S.}
\newblock \bibinfo{title}{Single-photon switch based on {R}ydberg blockade}.
\newblock \emph{\bibinfo{journal}{Phys. Rev. Lett.}}
  \textbf{\bibinfo{volume}{112}}, \bibinfo{pages}{073901}
  (\bibinfo{year}{2014}).

\bibitem{Naraschewski1999}
\bibinfo{author}{Naraschewski, M.} \& \bibinfo{author}{Glauber, R.~J.}
\newblock \bibinfo{title}{Spatial coherence and density correlations of trapped
  {B}ose gases}.
\newblock \emph{\bibinfo{journal}{Phys. Rev. A}} \textbf{\bibinfo{volume}{59}},
  \bibinfo{pages}{6} (\bibinfo{year}{1999}).

\bibitem{Burt1997}
\bibinfo{author}{Burt, E.~A.} \emph{et~al.}
\newblock \bibinfo{title}{Coherence, correlations, and collisions: What one
  learns about {B}ose-{E}instein condensates from their decay}.
\newblock \emph{\bibinfo{journal}{Phys. Rev. Lett.}}
  \textbf{\bibinfo{volume}{79}} (\bibinfo{year}{1997}).

\bibitem{Hodgman2011}
\bibinfo{author}{Hodgman, S.~S.}, \bibinfo{author}{Dall, R.~G.},
  \bibinfo{author}{Manning, A.~G.}, \bibinfo{author}{Baldwin, K. G.~H.} \&
  \bibinfo{author}{Truscott, A.~G.}
\newblock \bibinfo{title}{Direct measurement of long-range third-order
  coherence in {B}ose-{E}instein condensates}.
\newblock \emph{\bibinfo{journal}{Science}} \textbf{\bibinfo{volume}{331}},
  \bibinfo{pages}{1046--1049} (\bibinfo{year}{2011}).

\end{thebibliography}

\end{document}